# Energy, radial and time distributions of the charged particles at the maximum of electromagnetic showers initiated by 5-1000 GeV electrons in Fe, W and Pb

S.P. Denisov, V.N. Goryachev

Institute for High Energy Physics of the National Research Centre «Kurchatov Institute»
1 Nauki sqr, Protvino, Moscow region, 142281 Russian Federation

**Abstract**

The results of calculations of the charged particles energy, radial and time distributions at the maximum of electromagnetic showers initiated by electrons with energies from 5 to 1000 GeV in Fe, W and Pb are presented. It is shown that the shapes of energy distributions weekly depend on the electron energy, radial distributions for different materials become close to each other if radius is expressed in g/cm$^2$ and the time spread of the shower particles is in the picosecond range. Analysis of the data obtained allows us to conclude that a high Z material placed in a high energy electron beam can be used as a source of short and intense bunches of ultarelativistic positrons and electrons with subpicosecond time spread.

## 1. Introduction

In 1970 A.A.Tyapkin proposed to use a simple detector consisting of a lead convertor and a scintillation or Cherenkov counter behind it for $e,\gamma$ energy measurements[1]. This idea was based on the Rossi approximation of the electromagnetic shower development[2]. As an optimal convertor thickness A.A.Tyapkin suggested to use the value $t_{max}$ corresponding to the maximum flux $N_{max}$ of the charged shower particles since $t_{max}$ weakly (logarithmically) de-pends on the energy $E_0$ of the initial $e,\gamma$ and $N_{max}$ is almost proportional to $E_0$. At present this type of detector is often referred to as shower maximum detector. Since 1970 a number of studies (see, for example, [3-14]) were performed to investigate shower maximum detectors characteristics like energy and space resolutions and e/hadron and $\gamma/\pi^0$ separations which strongly depend on $N_{max}$ fluctuations and space and energy distributions of the charged shower particles.

In our article[15] distributions of charge particles multiplicities at $t_{max}$ are considered. In this report the results of calculations of the charged particles energy, radial and time distributions in Pb, W and Fe are presented. The calculations are based on GEANT4 10.01.p02 (Physical list FTFP_BERT) [16] with 700 micron range cut for all materials. Corresponding energy thresholds for $e^+$ and $e^-$ are close to 1 MeV for Pb and Fe and 1.6 MeV for W. Increase or decrease the range cut by a factor of two does not affect $N_{max}$ value within 0.5% statistical uncertainty[15] since the energy thresholds are much less than the average particle energy of ~50 MeV (see below). Note that the charged particles flux at $t_{max}$ consists mainly of $e^+$, $e^-$. For example··· the admixture of other particles at $E_0$=200 GeV is 0.02% only[15]. Diameter of all converters studied is 70 cm.

## 2. Energy distributions

Differential energy distributions of the charged particles at $t_{max}$ are shown in Fig.1. Their shapes weakly depend on the electron energy $E_0$. The same conclusion follows from integrated distributions presented in Tables 1-3. For example, the fraction of particles with energy below 50 MeV is varied from 0.742 to 0.752 (Fe), from 0.792 to 0.806 (W) and from 0.799 to 0.810 (Pb) when $E_0$ is changed from 5 to 1000 GeV.

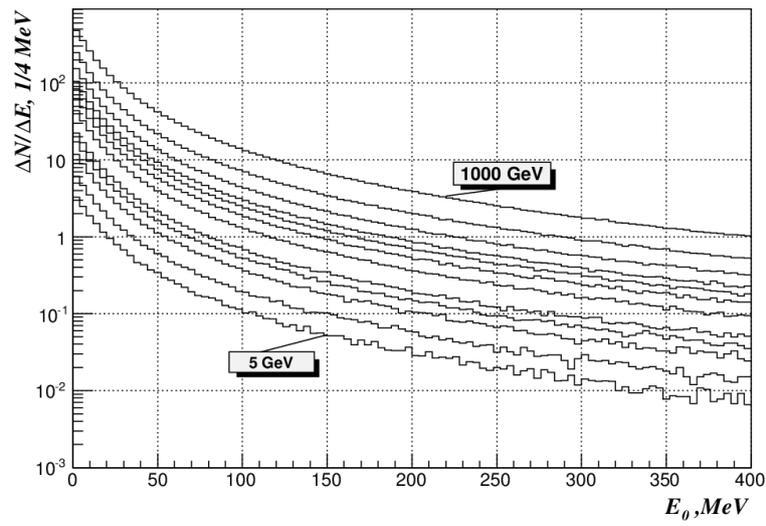

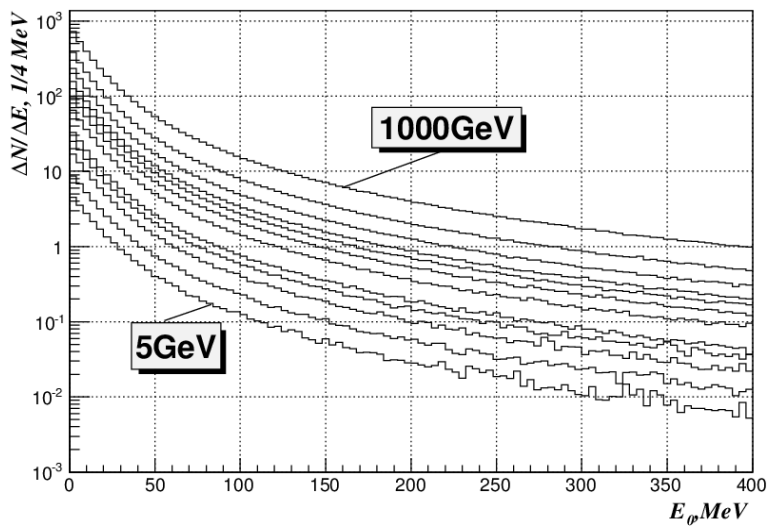

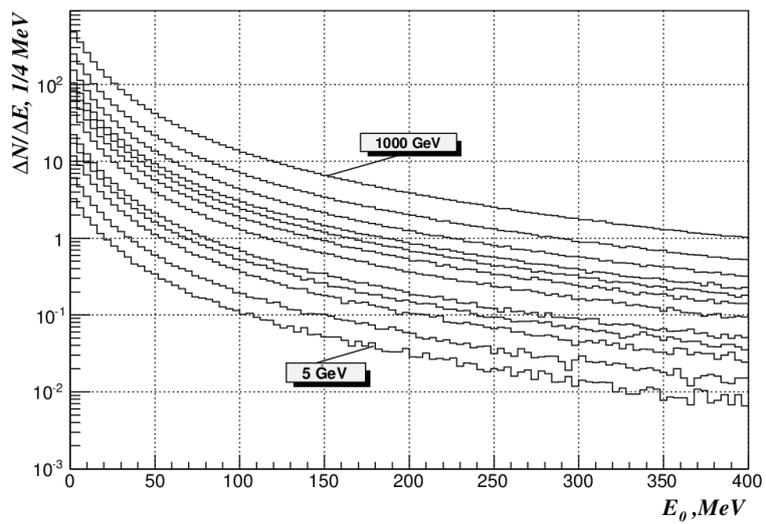

Fig. 1. Energy distributions of the charge particles at $t_{max}$.

Table 1. Fraction of particles with kinetic energy below $E$ at $t_{max}$ in Fe.

| $E_0$, GeV | $E$, MeV | | | | | | | | | |
|---|---|---|---|---|---|---|---|---|---|---|
| | 5 | 10 | 20 | 30 | 50 | 100 | 200 | 300 | 500 | 1000 |
| 5 | 0.199 | 0.341 | 0.519 | 0.624 | 0.742 | 0.864 | 0.938 | 0.964 | 0.984 | 0.996 |
| 10 | 0.202 | 0.344 | 0.522 | 0.626 | 0.744 | 0.863 | 0.936 | 0.961 | 0.981 | 0.994 |
| 20 | 0.206 | 0.350 | 0.525 | 0.629 | 0.745 | 0.863 | 0.934 | 0.960 | 0.980 | 0.993 |
| 30 | 0.207 | 0.351 | 0.528 | 0.631 | 0.747 | 0.863 | 0.934 | 0.959 | 0.979 | 0.992 |
| 40 | 0.207 | 0.352 | 0.529 | 0.632 | 0.746 | 0.863 | 0.933 | 0.958 | 0.978 | 0.992 |
| 80 | 0.210 | 0.355 | 0.531 | 0.634 | 0.748 | 0.863 | 0.932 | 0.957 | 0.977 | 0.991 |
| 120 | 0.211 | 0.356 | 0.532 | 0.634 | 0.748 | 0.862 | 0.932 | 0.957 | 0.977 | 0.991 |
| 160 | 0.212 | 0.357 | 0.534 | 0.636 | 0.750 | 0.863 | 0.932 | 0.957 | 0.977 | 0.991 |
| 200 | 0.213 | 0.359 | 0.535 | 0.638 | 0.751 | 0.864 | 0.932 | 0.957 | 0.977 | 0.991 |
| 300 | 0.214 | 0.361 | 0.537 | 0.639 | 0.752 | 0.864 | 0.932 | 0.956 | 0.976 | 0.990 |
| 500 | 0.215 | 0.362 | 0.539 | 0.640 | 0.752 | 0.864 | 0.932 | 0.956 | 0.976 | 0.990 |
| 1000 | 0.216 | 0.363 | 0.539 | 0.640 | 0.752 | 0.863 | 0.931 | 0.955 | 0.975 | 0.989 |

Table 2. Fraction of particles with kinetic energy below $E$ at $t_{max}$ in W.

| $E_0$, GeV | $E$, MeV | | | | | | | | | |
|---|---|---|---|---|---|---|---|---|---|---|
| | 5 | 10 | 20 | 30 | 50 | 100 | 200 | 300 | 500 | 1000 |
| 5 | 0.210 | 0.364 | 0.561 | 0.673 | 0.792 | 0.901 | 0.959 | 0.978 | 0.991 | 0.998 |
| 10 | 0.214 | 0.371 | 0.568 | 0.679 | 0.794 | 0.900 | 0.957 | 0.976 | 0.989 | 0.997 |
| 20 | 0.222 | 0.381 | 0.577 | 0.688 | 0.801 | 0.903 | 0.958 | 0.976 | 0.988 | 0.996 |
| 30 | 0.222 | 0.381 | 0.576 | 0.686 | 0.799 | 0.901 | 0.956 | 0.974 | 0.987 | 0.995 |
| 40 | 0.224 | 0.384 | 0.580 | 0.689 | 0.801 | 0.902 | 0.956 | 0.974 | 0.987 | 0.995 |
| 80 | 0.229 | 0.388 | 0.584 | 0.691 | 0.802 | 0.901 | 0.955 | 0.973 | 0.986 | 0.995 |
| 120 | 0.230 | 0.390 | 0.586 | 0.694 | 0.804 | 0.902 | 0.955 | 0.972 | 0.986 | 0.994 |
| 160 | 0.231 | 0.391 | 0.586 | 0.693 | 0.803 | 0.901 | 0.954 | 0.972 | 0.985 | 0.994 |
| 200 | 0.231 | 0.391 | 0.586 | 0.693 | 0.803 | 0.901 | 0.954 | 0.972 | 0.985 | 0.994 |
| 300 | 0.234 | 0.395 | 0.590 | 0.697 | 0.805 | 0.902 | 0.955 | 0.972 | 0.985 | 0.994 |
| 500 | 0.236 | 0.397 | 0.592 | 0.698 | 0.806 | 0.902 | 0.954 | 0.972 | 0.985 | 0.994 |
| 1000 | 0.234 | 0.395 | 0.588 | 0.694 | 0.802 | 0.899 | 0.952 | 0.970 | 0.983 | 0.993 |

Table 3. Fraction of particles with kinetic energy below $E$ at $t_{max}$ in Pb.

| $E_0$ (GeV) | $E$ (MeV) | | | | | | | | | |
|---|---|---|---|---|---|---|---|---|---|---|
| | 5 | 10 | 20 | 30 | 50 | 100 | 200 | 300 | 500 | 1000 |
| 5 | 0.217 | 0.373 | 0.570 | 0.682 | 0.799 | 0.905 | 0.961 | 0.978 | 0.991 | 0.998 |
| 10 | 0.225 | 0.381 | 0.577 | 0.687 | 0.802 | 0.904 | 0.960 | 0.977 | 0.989 | 0.997 |
| 20 | 0.229 | 0.385 | 0.580 | 0.690 | 0.802 | 0.904 | 0.958 | 0.976 | 0.988 | 0.996 |
| 30 | 0.230 | 0.387 | 0.582 | 0.691 | 0.803 | 0.903 | 0.957 | 0.975 | 0.988 | 0.996 |
| 40 | 0.233 | 0.390 | 0.585 | 0.693 | 0.803 | 0.903 | 0.957 | 0.974 | 0.987 | 0.996 |
| 80 | 0.237 | 0.396 | 0.589 | 0.696 | 0.806 | 0.904 | 0.957 | 0.974 | 0.987 | 0.995 |
| 120 | 0.238 | 0.396 | 0.590 | 0.696 | 0.805 | 0.903 | 0.956 | 0.973 | 0.986 | 0.995 |
| 160 | 0.240 | 0.399 | 0.593 | 0.699 | 0.808 | 0.905 | 0.957 | 0.974 | 0.986 | 0.995 |
| 200 | 0.240 | 0.400 | 0.592 | 0.699 | 0.807 | 0.904 | 0.956 | 0.973 | 0.986 | 0.995 |
| 300 | 0.243 | 0.403 | 0.595 | 0.701 | 0.809 | 0.905 | 0.956 | 0.973 | 0.986 | 0.995 |
| 500 | 0.244 | 0.404 | 0.597 | 0.702 | 0.810 | 0.905 | 0.956 | 0.973 | 0.986 | 0.994 |
| 1000 | 0.246 | 0.406 | 0.598 | 0.703 | 0.810 | 0.904 | 0.955 | 0.972 | 0.985 | 0.994 |

Fig. 2 demonstrates the dependencies of the average particle energy $<E>$ vs $E_0$. They are fitted to the formula

$$<E> = f_1 \cdot lnE_0 + f_2 \qquad (1)$$

where $f_1$ and $f_2$ are free parameters (see Table 4).

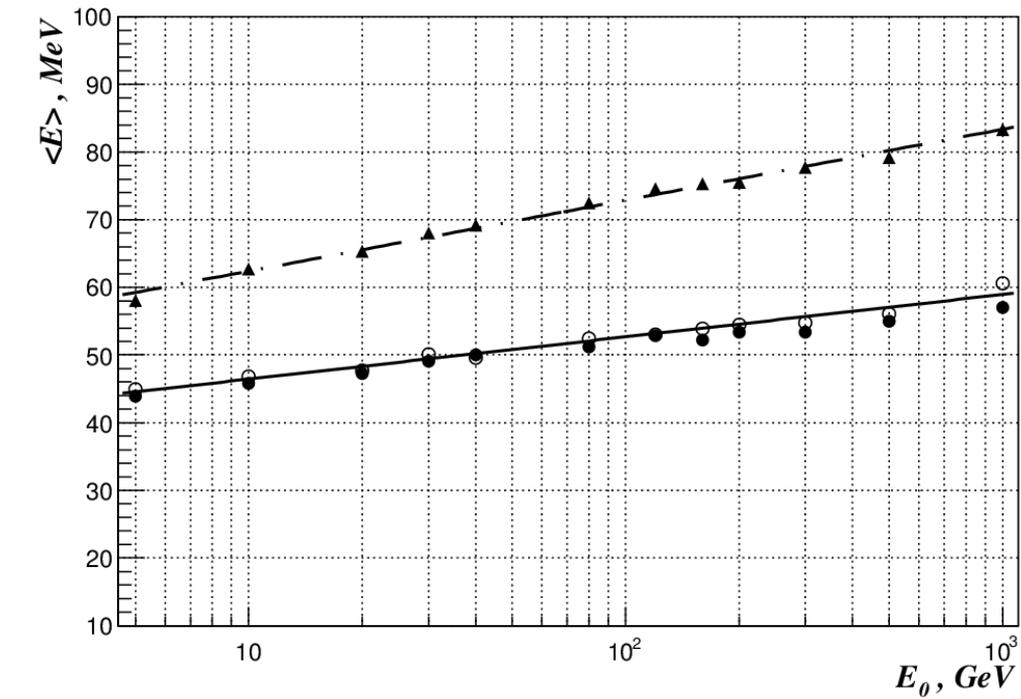

Fig. 2. Average particle energy at $t_{max}$ for Fe (▲), W (○) and Pb (●). The solid and dash-dotted lines represent the fit to equation (1) for W and Fe with parameters shown in Table 4. The results for W and Pb almost coincide with each other.

Table 4. Values of $f_1$ and $f_2$ parameters in the formula (1), $E_0$ is in GeV.

| Material | Fe | W | Pb |
|---|---|---|---|
| $f_1$, GeV | 4.55±0.13 | 2.71±0.15 | 2.39±0.10 |
| $f_2$, GeV | 51.92±0.61 | 40.19±0.68 | 40.52±0.47 |

### 3. Radial distributions

Radial distributions of the charge particles at $t_{max}$ are shown in Figs. 3,4 and Tables 5-7. They are rather narrow: a circle with a radius of 1 mm contains from 66% to 59% (W) and from 48% to 44% (Pb) of particles in the 5 to 1000 GeV energy range.

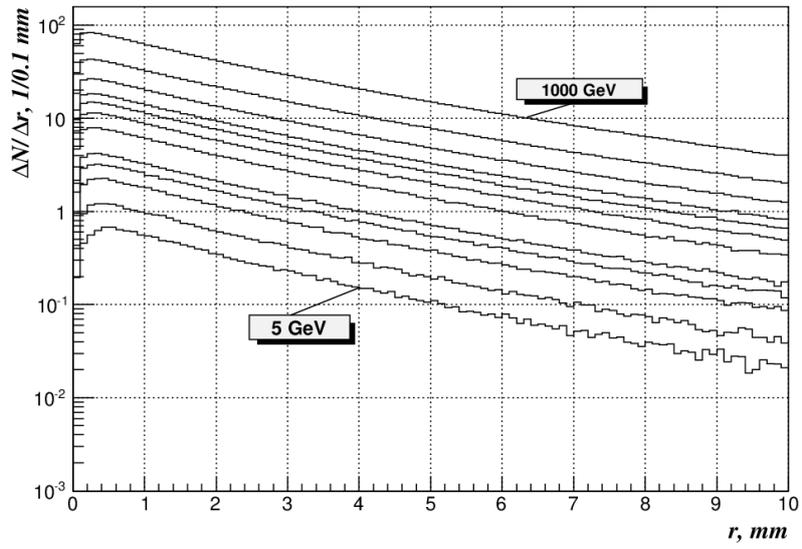
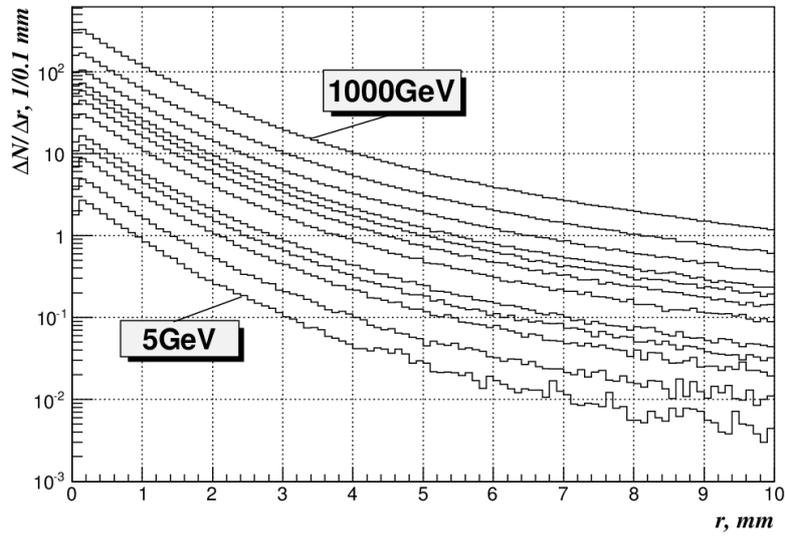
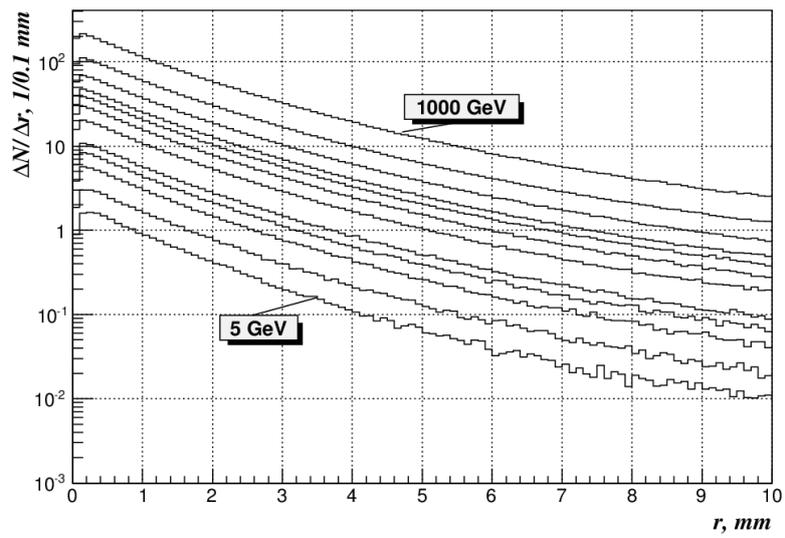

Fig. 3. Radial distributions of the charge particles at $t_{max}$.

Table 5. Fraction of particles inside a circle of radius $r$ at $t_{max}$ in Fe.

| $E_0$, GeV | $r$, mm | | | | | | | | | |
|---|---|---|---|---|---|---|---|---|---|---|
| | 0.5 | 1 | 2 | 3 | 5 | 10 | 20 | 30 | 50 | 100 |
| 5 | 0.127 | 0.284 | 0.511 | 0.652 | 0.812 | 0.940 | 0.980 | 0.988 | 0.993 | 0.997 |
| 10 | 0.136 | 0.285 | 0.501 | 0.641 | 0.801 | 0.933 | 0.977 | 0.986 | 0.992 | 0.997 |
| 20 | 0.139 | 0.283 | 0.494 | 0.631 | 0.789 | 0.926 | 0.976 | 0.986 | 0.992 | 0.997 |
| 30 | 0.140 | 0.282 | 0.488 | 0.625 | 0.784 | 0.923 | 0.974 | 0.985 | 0.992 | 0.997 |
| 40 | 0.142 | 0.281 | 0.485 | 0.621 | 0.780 | 0.920 | 0.973 | 0.985 | 0.992 | 0.997 |
| 80 | 0.143 | 0.280 | 0.480 | 0.614 | 0.772 | 0.916 | 0.971 | 0.983 | 0.991 | 0.997 |
| 120 | 0.144 | 0.280 | 0.479 | 0.612 | 0.770 | 0.913 | 0.970 | 0.983 | 0.991 | 0.997 |
| 160 | 0.143 | 0.277 | 0.474 | 0.608 | 0.766 | 0.911 | 0.969 | 0.982 | 0.991 | 0.997 |
| 200 | 0.143 | 0.276 | 0.472 | 0.605 | 0.763 | 0.909 | 0.968 | 0.982 | 0.991 | 0.997 |
| 300 | 0.142 | 0.274 | 0.468 | 0.600 | 0.758 | 0.906 | 0.967 | 0.981 | 0.991 | 0.997 |
| 500 | 0.143 | 0.274 | 0.466 | 0.597 | 0.755 | 0.903 | 0.966 | 0.981 | 0.990 | 0.997 |
| 1000 | 0.145 | 0.275 | 0.466 | 0.596 | 0.753 | 0.901 | 0.965 | 0.980 | 0.990 | 0.996 |

Table 6. Fraction of particles inside a circle of radius $r$ at $t_{max}$ in W.

| $E_0$, GeV | $r$, mm | | | | | | | | | |
|---|---|---|---|---|---|---|---|---|---|---|
| | 0.5 | 1 | 2 | 3 | 5 | 10 | 20 | 30 | 50 | 100 |
| 5 | 0.417 | 0.655 | 0.851 | 0.919 | 0.961 | 0.982 | 0.990 | 0.992 | 0.995 | 0.998 |
| 10 | 0.409 | 0.641 | 0.841 | 0.912 | 0.957 | 0.981 | 0.989 | 0.992 | 0.994 | 0.997 |
| 20 | 0.392 | 0.620 | 0.823 | 0.899 | 0.951 | 0.978 | 0.988 | 0.991 | 0.994 | 0.997 |
| 30 | 0.395 | 0.620 | 0.822 | 0.898 | 0.950 | 0.978 | 0.988 | 0.991 | 0.993 | 0.997 |
| 40 | 0.390 | 0.614 | 0.817 | 0.894 | 0.948 | 0.977 | 0.988 | 0.991 | 0.993 | 0.997 |
| 80 | 0.385 | 0.606 | 0.809 | 0.888 | 0.944 | 0.975 | 0.987 | 0.990 | 0.993 | 0.996 |
| 120 | 0.382 | 0.601 | 0.804 | 0.885 | 0.942 | 0.974 | 0.986 | 0.990 | 0.993 | 0.996 |
| 160 | 0.383 | 0.601 | 0.803 | 0.883 | 0.941 | 0.974 | 0.986 | 0.989 | 0.992 | 0.996 |
| 200 | 0.382 | 0.600 | 0.802 | 0.882 | 0.940 | 0.974 | 0.986 | 0.989 | 0.992 | 0.996 |
| 300 | 0.377 | 0.593 | 0.796 | 0.878 | 0.937 | 0.972 | 0.986 | 0.989 | 0.992 | 0.996 |
| 500 | 0.375 | 0.589 | 0.792 | 0.874 | 0.935 | 0.972 | 0.985 | 0.989 | 0.992 | 0.996 |
| 1000 | 0.375 | 0.594 | 0.794 | 0.875 | 0.935 | 0.971 | 0.985 | 0.989 | 0.992 | 0.996 |

Table 7. Fraction of particles inside a circle of radius $r$ at $t_{max}$ in Pb.

| $E_e$, GeV | $r$, mm | | | | | | | | | |
|---|---|---|---|---|---|---|---|---|---|---|
| | 0.5 | 1 | 2 | 3 | 5 | 10 | 20 | 30 | 50 | 100 |
| 5 | 0.270 | 0.481 | 0.718 | 0.830 | 0.920 | 0.969 | 0.985 | 0.989 | 0.993 | 0.997 |
| 10 | 0.268 | 0.473 | 0.704 | 0.818 | 0.912 | 0.966 | 0.984 | 0.989 | 0.992 | 0.997 |
| 20 | 0.264 | 0.463 | 0.691 | 0.805 | 0.903 | 0.962 | 0.982 | 0.988 | 0.992 | 0.997 |
| 30 | 0.266 | 0.462 | 0.687 | 0.801 | 0.899 | 0.961 | 0.982 | 0.987 | 0.991 | 0.996 |
| 40 | 0.264 | 0.458 | 0.683 | 0.797 | 0.897 | 0.959 | 0.981 | 0.987 | 0.992 | 0.996 |
| 80 | 0.260 | 0.450 | 0.673 | 0.788 | 0.890 | 0.956 | 0.980 | 0.986 | 0.991 | 0.996 |
| 120 | 0.260 | 0.449 | 0.670 | 0.785 | 0.888 | 0.955 | 0.979 | 0.986 | 0.991 | 0.996 |
| 160 | 0.257 | 0.444 | 0.665 | 0.780 | 0.884 | 0.953 | 0.979 | 0.986 | 0.991 | 0.996 |
| 200 | 0.257 | 0.444 | 0.664 | 0.780 | 0.884 | 0.953 | 0.979 | 0.986 | 0.991 | 0.996 |
| 300 | 0.254 | 0.440 | 0.659 | 0.775 | 0.880 | 0.951 | 0.978 | 0.985 | 0.991 | 0.996 |
| 500 | 0.254 | 0.437 | 0.655 | 0.771 | 0.877 | 0.949 | 0.977 | 0.985 | 0.990 | 0.996 |
| 1000 | 0.254 | 0.436 | 0.652 | 0.767 | 0.874 | 0.948 | 0.977 | 0.985 | 0.990 | 0.995 |

Radial distributions for different materials become close to each other if radius is expressed in $g/cm^2$ (see Fig.4). The following formula was used to fit the integral distributions shown in Fig.4:

$$f(r) = 1 - f_0 \cdot e^{-sr} - (1 - f_0) \cdot e^{-t \cdot r}, \tag{2}$$

where $f_0$, $s$ and $t$ are free parameters (see Table 8).

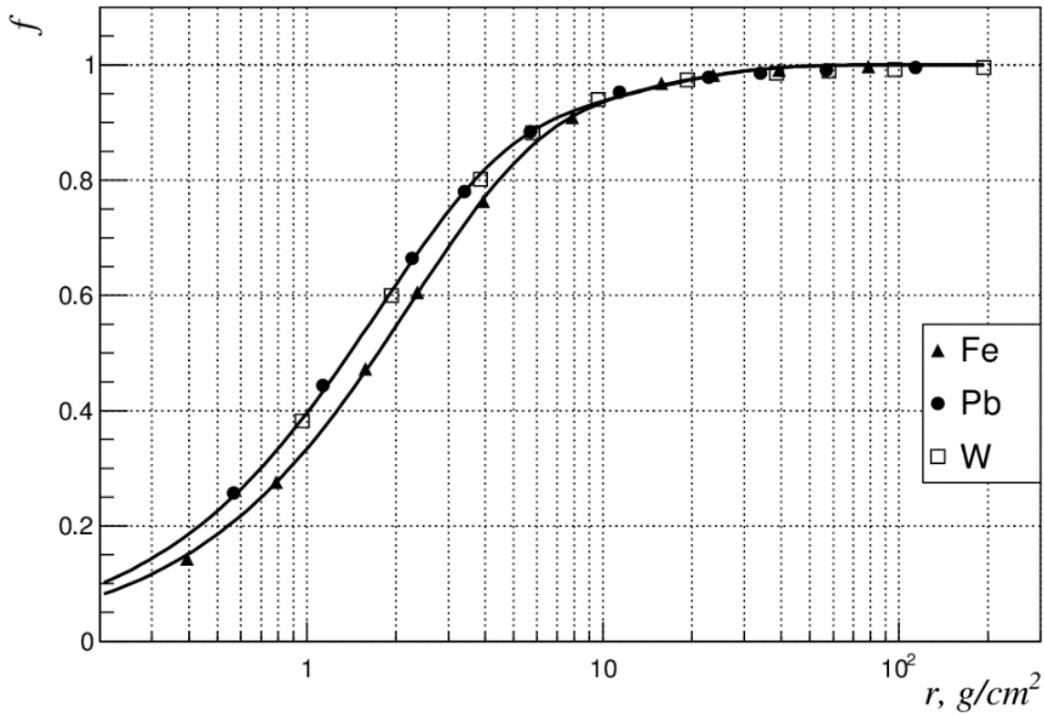

Fig. 4. Fraction of particles inside the ring of radius $r$ for $E_0$=200 GeV. The curves present the fit to eq. (2) for W and Pb (top) and Fe (bottom) with parameters shown in Table 8.

Table 8. Parameters values in the formula (2), $r$ is in $g/cm^2$.

| Material | $f_0$ | $s$ | $t$ |
|---|---|---|---|
| Fe | 0.13 | 0.085 | 0.47 |
| Pb, W | 0.14 | 0.086 | 0.59 |

## 4. Time distributions

Time distributions of the charged particles for 200 GeV showers in W are shown in Fig. 5. They are extremely narrow: 90% of particles are in the time window of 4 ps. For the 90% of particles inside a circle of $r = 1$mm the time spread is equal to 0.8 ps.

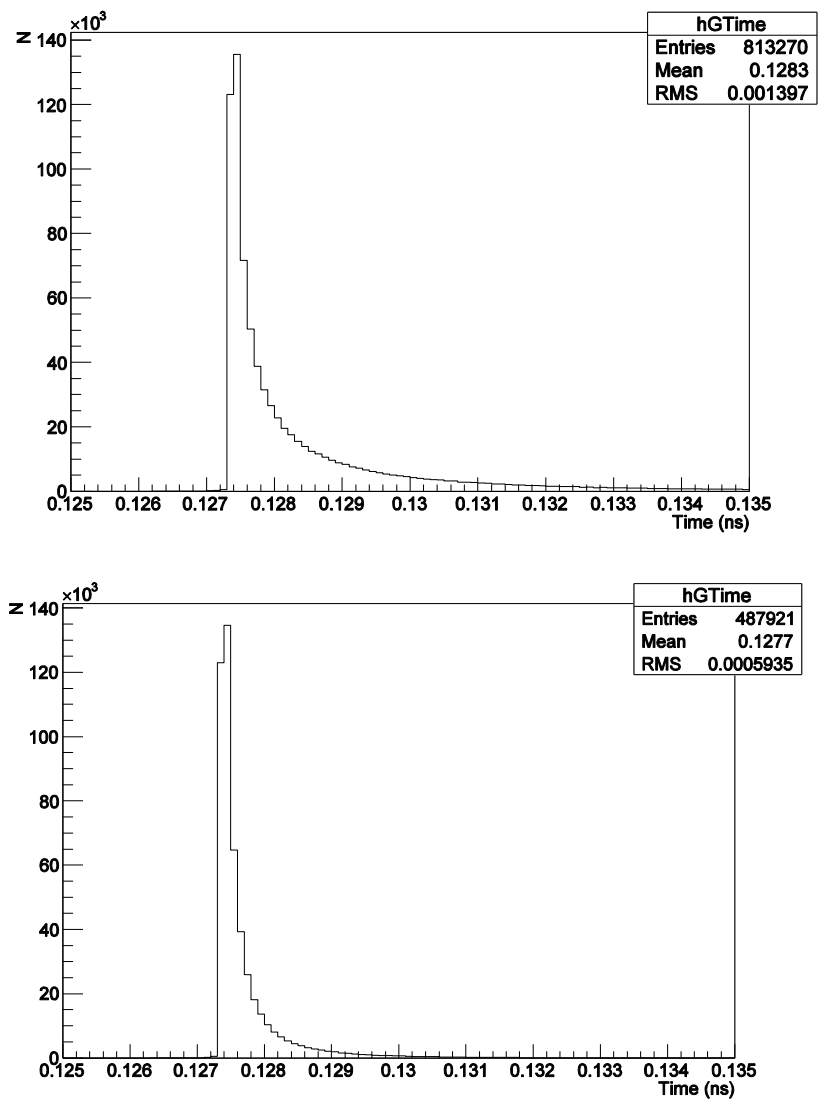

Fig.5. Time distributions for all charged particles (top) and those inside 1 mm circle (bottom) at $t_{max}$ for W converter and $E_0$=200 GeV.

## 5. Conclusions

Calculations of the energy, radial and time distributions of the charge particles at the maximum of electromagnetic showers initiated by 5 to 1000 GeV electrons in Fe, W, and Pb are performed using GEANT4. It iss shown that the shapes of energy distributions and the average particle energy weakly depend on the incident particle energy and radial distributions for different materials become close to each other if radius is expressed in $g/cm^2$. The radial and time distributions appeared to be narrow. For example, 200 GeV incident electron produces at $t_{max}$ in W about $5 \cdot 10^2$ particles within the radius r<1 mm. Their time spread is 0.8 ps and average energy is 82 MeV. Thus a high $Z$ converter with thickness of $t_{max}$ placed in a high energy electron beam can be used as a source of short and intense bunches of ultarelativistic positrons and electrons with subpicosecond time spread.

## Acknowledgment

We gratefully acknowledge the help of D.S.Denisov, T.Z.Gurova, A.V.Kozelov and D.A.Stoyanova in preparation of this manuscript. This work was supported in part by the Russian Foundation for Basic Research under grant #17-02-00120.